# Biophysical Measurements of Cells, Microtubules, and DNA with an Atomic Force Microscope

Luka M. Devenica, Clay Contee, Raysa Cabrejo, and Matthew Kurek

*Department of Physics, Amherst College, Amherst, MA 01002*

Edward F. Deveney

*Department of Physics, Bridgewater State University, Bridgewater, MA 02325*

Ashley R. Carter[†]

*Department of Physics, Amherst College, Amherst, MA 01002*

(Dated: December 14, 2014)

## Abstract

Atomic force microscopes (AFMs) are ubiquitous in research laboratories and have recently been priced for use in teaching laboratories. Here we review several AFM platforms (Dimension 3000 by Digital Instruments, EasyScan2 by Nanosurf, ezAFM by Nanomagnetics, and TKAFM by Thorlabs) and describe various biophysical experiments that could be done in the teaching laboratory using these instruments. In particular, we focus on experiments that image biological materials and quantify biophysical parameters: 1) imaging cells to determine membrane tension, 2) imaging microtubules to determine their persistence length, 3) imaging the random walk of DNA molecules to determine their contour length, and 4) imaging stretched DNA molecules to measure the tensional force.

## I. INTRODUCTION

Today, we rely on computing devices that have manufactured chips with specific atomic-scale properties. However, in the early 1980's atomic-scale surface science and materials research was just underway at IBM. The computing giant was investing heavily in basic research and hired a team of researchers: Gerd Binnig, Heinrich Rohrer, Christoph Gerber, and Edi Weibel, to perform local spectroscopy of surfaces using the elusive technique of electron tunneling.[1] After demonstrating that electrons could tunnel from a sharp, conducting probe through a vacuum to a nearby metal surface, they began to use the conducting probe to map the local properties.[2,3] Scanning this conducting probe revealed incredibly sharp images of the surface, leading to the discovery of the scanning tunneling microscope (STM) and the 1986 Nobel Prize for Binnig and Rohrer.

However, the STM was only able to measure conductive samples. To measure the surface properties of insulators, a new technique would need to be developed. Thus, Binnig and Gerber collaborated with another physicist, Calvin Quate of Stanford, to develop an atomic force microscope (AFM).[4] In the AFM, a sharp probe scans across the surface and moves up and down with the changing topography (Fig. 1A). A laser beam reflecting off the back of the probe, instead of the tunneling current, is used to detect atomic-scale motion in height. This allows for imaging of any type of surface with atomic resolution, including biological materials.[5] Today, the AFM and STM are instrumental in semiconductor manufacturing where patterning and visualization of single atoms is required.

Recently, a number of companies have started to offer AFMs priced for educational use. We review several of these instruments (the Dimension 3000 by Digital Instruments, the EasyScan2 by Nanosurf, the ezAFM by Nanomagnetics, and the TKAFM by Thorlabs) and find a wide range of axial resolutions (0.1-100 nm) and capabilities. With the introduction of these AFMs, there is now a need for laboratory curricula to train the next generation of scientists. While some laboratory coursework has been developed,[6-13] we are particularly interested in developing AFM laboratories that cover biophysics concepts.

AFM has become exceedingly popular in biophysics research because it allows scientists to manipulate and image proteins, nucleic acids, membranes, or cells in ambient conditions in liquid.[14,15] Here we describe four biophysical laboratories that give students the opportunity to image biological samples and quantify biophysical parameters even on educational AFMs that lack atomic-scale precision. In the first laboratory, we image budding yeast cells and determine the relative mother-bud membrane tension. This laboratory is particularly designed for an AFM with a higher noise floor (>20 nm). For an AFM with a lower noise floor (1-10 nm), we image microtubules fixed to the surface with different concentrations of glutaraldehyde and measure the apparent persistence length. For AFMs with atomic resolution (~0.1 nm), we image the random walk of individual DNA molecules and measure their contour length. Finally, AFMs with nanometer resolution can image DNA molecules that have been stretched with a flow, allowing students to determine the tensional force. All laboratories can be completed in a three hour time slot, perfect for implementation in a biophysics or modern physics course.

## II. EXPERIMENTAL MATERIALS AND METHODS

We begin by testing four AFM platforms with slightly different setups. The goal in reviewing this large range of educational AFMs is to develop a series of biophysical laboratories that can be used with this wide assortment of instruments.

In a typical AFM, a sharp probe (radius of ~10 nm) at the end of a long cantilever (length of ~100 μm) scans along the sample surface (Fig. 1A). To detect the position of the probe, a laser is reflected off the backside of the cantilever onto a quadrant photodiode (QPD). When the probe encounters a change in the height of the surface, the cantilever angle changes, and the laser deflects to a new location on the QPD, recording the change in height ($z$). A single scan in one dimension (*e.g.* $x$) will record a profile of the surface ($z(x)$). To create a two dimensional image, the AFM alternates between scanning the probe in $x$ and incrementing the probe in $y$. Three of the AFMs we tested operate in this manner (the Dimension 3000, the EasyScan2, and the ezAFM).

In the TKAFM,[16] the deflection of the cantilever is detected by a Fabry-Perot interferometer (Fig. 1B). The interferometer consists of a laser coupled into a fiber with a 50:50 splitter (*not shown*). One of the outputs of the fiber is unused and the other end is placed within a millimeter or so of the backside of the cantilever. The laser light reflected from the cantilever and the light reflected from the fiber-air interface travel back through the fiber and the 50:50 splitter and interfere at a photodiode. When the probe encounters a feature on the surface, the cantilever-fiber distance decreases, and the interference pattern on the photodiode changes. Scanning the probe in *x* and *y* creates an image of the topography of the surface.

Typically, AFMs operate in either contact mode (EasyScan2 and TKAFM) or tapping mode (Dimension 3000 and ezAFM). In contact mode, the probe is scanned along the surface at a fixed height or a fixed relative height to the sample. In tapping mode, the probe-sample distance is modulated such that the probe only taps the sample at the peak of the modulation, minimizing tip-sample interaction. This decreases the frictional forces and limits sample and tip damage.[15] Soft biological samples are typically imaged in tapping mode. The exact configuration for each system is listed below.

### A. Dimension 3000 AFM

The Dimension 3000 AFM by Bruker Nano/Veeco Metrology/Digital Instruments is a research grade system that was first available in the 1990's and can now be purchased second hand. Our system configuration uses a Nanoscope IIIa controller running the Digital Instruments software, as well as the Dimension 3000 scan head, base, sample plate, and probe holder (tapping mode, price ~$65k from Advanced Surface Microscopy, Inc., 2014). The scan head and sample plate are not enclosed and are located on a 3/16"-thick TMC breadboard for passive vibration isolation.

### B. EasyScan2

The EasyScan2 by Nanosurf is designed as a portable, educational instrument that can be outfitted to fulfill several roles in the laboratory. Instruments can be purchased with an STM or AFM head, and the AFM head can be outfitted for contact mode, tapping mode, or both. Our system configuration includes a 70 by 70 µm scanner and EasyScan2 software (contact mode, price ~$27k, 2007). This system has recently been replaced by the Naio model (contact mode, price ~$25k, 2015). The AFM probe on the EasyScan2 is automatically aligned to the laser using an alignment grid system on the AFM chip. Thus, probes must meet the following two conditions: 1) the probe chip must contain alignment grooves and 2) the cantilever must be 225 µm long. There is also a line of probes from NanoSensors (XY Alignment Compatible) that are compatible with the instrument and have varying cantilever lengths. The EasyScan2 was placed on a 2" optical breadboard with vibration isolators (Newport, VH3648W-OPT). A compressor (Rolair Systems, 2.3 gallon, 125 psi) floated the table at 92 psi.

### C. ezAFM

The ezAFM by Nanomagnetics is designed as a portable system for educational use. Our system configuration includes a 40 by 40 µm scanner and the ezAFM software (tapping mode, price $15k, 2014). The scan head and sample are enclosed in a cylindrical metal case that fits in the palm of your hand. The controller also has a small profile and the whole thing can be stored in a briefcase for easy travel. The ezAFM, like the EasyScan2, requires AFM probes to have the alignment grid system, which facilitates changing the probe. A computer runs the controller with the ezAFM v3.29 software. The system comes with an inexpensive passive vibration isolator, but we used a small vibration isolation table (BM-10 from Minus-K, $2.5K) to increase performance without sacrificing portability. For higher resolution experiments, we couple the BM-10 with a 4 by 8 foot floating optics table (CleanTop with Gimbal PistonTM isolators from TMC, $10k).

### D. TKAFM

The TKAFM[16] by Thorlabs is a modular teaching kit that allows students to build their own AFM over several laboratory periods. Our system configuration includes a beta version of the TKAFM mounted on a ½" optical breadboard along with a beta version of the TKAFM software (contact mode, price ~$8k, 2012). The TKAFM and TKAFM software are currently being updated and a full release of the product is forthcoming. The beta version of the TKAFM is delivered as a working contact mode instrument that can then be taken apart before the laboratory begins. The sample is scanned using a 3D piezo-electric stage (NanoMax, Thorlabs). The AFM probe is attached to a 1D piezo-electric stage operating in the axial dimension, and probe height is detected by a Fabry-Perot interferometer system (Fig. 1B, $\lambda = 625$ nm, 5 mW). To position the fiber, the system comes with a 6-axis kinematic mount. We added an additional tip-tilt stage (APR001, Thorlabs) underneath the sample so that our images would not require flattening. An angle bracket (ABS002, Thorlabs) increased the height of the AFM probe above the surface to accommodate the tip-tilt stage. Optics were placed on a 4 by 8 foot floating optics table (CleanTop with Gimbal PistonTM isolators from TMC) without an enclosure. The beta version of the TKAFM software does not calculate height and instead lists values of pixels as an

8-bit number. We convert this pixel value to nanometers assuming that a white pixel (value = 255) is 100 nm taller than a black pixel (value = 0). This rough calibration is based on a linear fit of our photodiode voltage versus height curve.

**E. AFM Probe Selection**

We selected AFM probes based on price, purpose, and compatibility with our AFM platforms. Individual probes are typically sold in packs of 10-50 at $25-$75 a probe, with lower prices corresponding to higher volume purchases. Thus, price is important when selecting probes even though only one probe is needed for a 3 hour laboratory.

The beta version of the TKAFM came with interdigitated cantilevers,[6] and we used these for all images taken with the instrument. We note that the beta version of the TKAFM that we use does not actually take advantage of the interdigitated cantilever design and that other probes could be used with the system.

For the EasyScan2 we used the PPP-XYNCSTR probe (NanoSensors, resonant frequency = 160 kHz, force constant = 7.4 N/m, length = 150 μm, tip radius < 7 nm). This probe has a high resonant frequency for fast scanning and is typically used with tapping mode instruments. However, we found the lower force constant to be useful when imaging in contact mode.

For the tapping mode instruments (Dimension 3000 and ezAFM) we used a variety of probes. For imaging hard samples with the Dimension 3000 and ezAFM, we selected the cone-shaped PPP-NCLR probe (NanoSensors, resonant frequency = 190 kHz, force constant = 48 N/m, length = 225 μm, tip radius < 10 nm). This probe was chosen due to its high force constant and high resonant frequency, which allowed for fast scanning. For imaging soft samples with the Dimension 3000 and ezAFM, our probe of choice was the PPP-XYNCSTR. However, at times we used the SSS-FMR probe (NanoSensors, resonant frequency = 75 kHz, force constant = 2.8 N/m, length = 225 μm, tip radius < 2 nm) with the ezAFM and the HiRes-C19/Au-Cr probe (MicroMasch, resonant frequency = 65 kHz, force constant = 0.5 N/m, length = 125 μm, tip radius < 1 nm) with the Dimension 3000. The smaller tip radius on these probes improved lateral resolution. When possible, probes were handled using cantilever tweezers (Ted-Pella) and an electric static discharge mat (NanoAndMore) to limit damage. Table 1 lists the AFM probes used in each image.

**F. Sample Preparation**

Our goal in choosing samples was to select samples that are typically used in biophysics research and would be easy to prepare during a laboratory class period. Three such samples are DNA, microtubules, and cells. We prepare all samples for imaging in air.

To prepare DNA samples, we first used double-sided scotch tape to adhere a mica coverslip (Ted-Pella, 10-mm-diameter) to a metal specimen disk (Nanomagnetics, 28-mm-diameter). The metal disk allows the sample to be magnetically attached to the ezAFM. (The other AFMs did not require a particular sample size or holder.) We then pressed a piece of single-sided scotch tape to the mica and quickly removed the tape, leaving an atomically flat, clean layer. This procedure was repeated 3-4 times and the mica coverslip was inspected with a 5X stereoscope to make sure there were minimal cracks. Next, we diluted double-stranded DNA from bacteriophage λ (New England Biolabs, $L_0$ = 48,502 base pairs or 16.4 µm, 500 µg/mL) to 1 µg/mL in a solution of 1 mM magnesium acetate (Sigma). (All salt solutions and buffers listed in this section are reagent grade and are filtered with a 0.2 µm filter.) We then immediately pipetted 20 µL of the DNA-magnesium acetate solution onto the mica coverslip and waited 5 min before rinsing with 500 µL of filtered deionized water. The positively charged magnesium coats the negatively charged mica surface and traps the negatively charged DNA in a salt layer.[17] If the surface is not rinsed properly before drying, the salt layer can build up, obscuring the DNA adhered to the surface.[17] During rinsing, excess water can be dripped into the sink or can be absorbed with filter paper, before air drying. To prepare samples with single-stranded DNA, viral DNA from M13mp18 (New England Biolabs, $L_0$ = 7249 base pairs or 2450 nm, 250 µg/mL) was diluted to 0.5 µg/mL and prepared using the same method as the double-stranded DNA. To prepare λ phage DNA samples in the presence of a flow, we pipetted 500 µL of the DNA-magnesium acetate solution onto the mica cover slip while holding the sample at an angle. We then waited 5 min and rinsed with deionized water as before. Total prep time is ~10 minutes and samples can be kept at room temperature for months. To make new samples, just remove the top few layers of mica with single-sided scotch tape and begin again.

To prepare samples with microtubules, we first prepared the mica coverslips as before. Then, we added 5 µL of 1 mg/mL poly-lysine (0.1%, molecular weight >300,000 u, Sigma) to the mica by spreading it on the surface with a rectangular glass cover slip before allowing the surface to air dry. The poly-lysine is positively charged and serves the same purpose as the magnesium acetate. However, the poly-lysine layer creates a surface roughness of about 3-5 nm peak-to-peak. Next, we diluted a stock of 5 mg/mL microtubules to 0.1 mg/mL in PEM buffer [made with 100 mM 1,4-Piperazinediethanesulfonic acid (PIPES, pH 7.2), 2 mM ethylene glycol tetraacetic acid (EGTA), and 1 mM magnesium sulfate ($MgSO_4$) from Sigma] and either 1% or 4% glutaraldehyde (Sigma) under a fume hood. Finally, we immediately added 10 µL of the solution to the poly-lysine coated mica and let sit for 5 min before rinsing with 1 mL of filtered deionized water and air drying. Taxol-stabilized microtubules were given to us by colleagues, but microtubule kits are available for purchase (Cytoskeleton, Inc). The glutaraldehyde and poly-lysine should be kept on ice during the procedure, while the microtubules should be kept at room temperature. Prep time is 15-20 minutes (not including buffer preparation) and samples will remain intact for several months or longer at room temperature.

To prepare cells, we first cultured *Saccharomyces cerevisiae* (baker's yeast) according to established protocols.[18] Briefly, cells were grown on yeast extract peptone dextrose (YPD), which is a complete medium for yeast growth, containing 5 g yeast extract, 10 g tryptone, 5 g sodium chloride (NaCl), 20 g bactoagar, and 1 L of water. Warm medium was poured into petri dishes and allowed to cool before streaking a line of yeast cells onto the plate with a sterilized toothpick. After 3-5 days of incubation at 30°C, a single colony was chosen and used to inoculate a 5 mL liquid culture (5 g yeast extract, 10 g tryptone, 5 g NaCl and 1 L of water). The liquid culture was incubated at 30 °C for 8 hours to obtain budding yeast and 24-48 hours to obtain yeast without buds. We then centrifuged 1.5 mL of the liquid culture and removed the YPD. We resuspended the cells in 50 µL of phosphate buffered saline (PBS). To image the cells, we coated a glass coverslip with 10 µL of poly-lysine and dried the coverslip on a warm hot plate. Next, we added 5 µL of cells onto the coverslip. Finally, we waited 5 min and rinsed the coverslip with 400 µL of water. Samples were kept at 4 °C for several weeks. Yeast cells were chosen for imaging because they have a relatively hard cell wall that can withstand the frictional forces of contact mode and because students can use the cells cultured by our introductory biology class.

In addition to these student-prepared samples, we also used two commercially available AFM standards. The AFM height standard (HS-20MG from Budget Sensors) has 20-nm-tall, silicon pillars that are 5 µm in pitch. Arrows on the sample direct the user to the fabricated area. The DNA standard (DNA01 from K-Tek) has linear DNA molecules that are 3000 base pairs or 1009 nm long adhered to the surface at a nominal concentration of 0.5-7 molecules/µm$^2$, though measured concentrations were typically 500 molecules/µm$^2$. DNA molecules were adhered to the mica with 3-aminopropyltriethoxysilane (APTES).[19] According to the manufacturer heights of these molecules should be half a nanometer at 3-5% relative humidity. Higher humidity may obscure the DNA.[17] Make sure to keep the sample enclosed in a plastic bag with a desiccant at room temperature.

### G. Data Acquisition

Data acquisition for the four instruments is relatively similar even though each instrument uses its own custom software. After the AFM probe is loaded and the sample is in place, we first align the laser to the probe (which is automatic for systems that use AFM probes with alignment grooves). For tapping mode systems, we also determine the resonant frequency of the probe. Next, the AFM probe is brought into contact with the surface by moving the tip axially toward the surface while monitoring the photodiode signal. This is automatic in all systems except the TKAFM. Finally, we enter the scan parameters into the software and start scanning in $x$ (typically the horizontal dimension) while incrementing in $y$ (typically the vertical dimension) after each scan. After scanning begins, we adjust the proportional, integral, and derivative gain settings. To initially set the gain, we either use the manufacturer's recommendations or tune the controller according to the Zeigler-Nichols tuning method.[20] This initial setting can then be modified by hand. We typically scan at a gain setting where the proportional gain is just below the setting that causes oscillation. Scanning parameters for all of the images in the document are listed in Table 1. For images taken with the TKAFM the raw data is stored. For images taken with the other three AFM systems we remove sample tilt. To do this with the ezAFM, we "plane flatten" and "equalize" in $x$ and $y$ using the ezAFM software. For images taken with the Dimension 3000 or EasyScan2 we export the raw data and use either "linewise leveling" or "global leveling" in SPIP v6.3.2 or "plane leveling" in Gwyddion.[21] Images are also cropped where indicated and the axial scale bar is set to an appropriate value. Changing the axial scale bar may cause some pixels to saturate in the displayed images; height measurements are only performed on unsaturated data. No other filtering or image modification is made.

### H. Data Analysis

Images are analyzed by measuring the lengths or heights of objects in the image. To do this, we manually fit a line or ellipse to the feature in the profile or image. Measurements are either made in the software that comes with each AFM or by analyzing the image in ImageJ[22] or IGOR Pro. Profiles are produced using the AFM software and analyzed in IGOR Pro.

### III. AFM COMPARISON

The four AFMs that we tested have a wide variety of imaging abilities that will limit what laboratories can be done with these instruments. One way to compare instruments directly is to compare the theoretical resolution of the scan head in $z$, which is 0.09 nm for the Dimension 3000,[23] 0.21 nm for the EasyScan2,[24] 0.002 nm for the ezAFM,[25] and 5 nm for the TKAFM.[26] This number represents the smallest distance the stage could theoretically be moved if other noise sources like vibration are sufficiently low. To get a sense of the dominant source of noise in our student laboratory, we first imaged 20-nm-tall, silicon pillars (Fig. 2) on our AFM height standard. The Dimension 3000, EasyScan2, ezAFM, and the TKAFM measured pillars that were $20.0 \pm 0.5$ nm, $20.0 \pm 0.6$ nm, $19.8 \pm 0.8$ nm, or $16 \pm 12$ nm tall, respectively. Height measurements are the average ± standard deviation for >10 pillars. We also measured the noise level for each image by taking the standard deviation for 25 pixels that were all at the same nominal height in one line scan. We repeated this measurement 5 times and quote the average ± standard deviation. The Dimension 3000, EasyScan2, ezAFM, and the TKAFM had a noise level of $0.40 \pm 0.06$ nm, $0.32 \pm 0.09$ nm, $0.45 \pm 0.04$ nm, or $16 \pm 4$ nm, respectively. We note that each image was taken with settings that allowed us to achieve the lowest noise floor for the different instruments (see Table 1 for the settings). Most notably, we decreased the lateral resolution of the TKAFM to optimize the height resolution.

From this data, we find that there is a clear difference in the noise level of the TKAFM in comparison to the other three educational instruments. One reason for this order of magnitude difference in the noise floor is the order of magnitude difference in the resolution of the scan head for the TKAFM, which decreases the price of the TKAFM by a factor of 2-10 from the other instruments. The other reason is that the instrument is a kit that is completely assembled by the students and is therefore subject to a large range of possible noise sources. For the best results, please see company literature. The beta version of the TKAFM that we test here will most likely be replaced by a more robust version.

In conducting this experiment, we found that the dominant source of noise was mechanical vibration. For example, if we remove the vibration isolation table from the ezAFM and use only the passive isolation sold with the instrument, the noise level we measure is 1.1 ± 0.1 nm, an increase of a factor of two. This indicates that care should be taken to isolate the instruments from vibration if an atomic-scale (~0.1 nm) noise floor is needed. For further noise reduction, active methods[27,28] could be employed. However, active noise reduction is not necessary for the experiments that we describe as the Dimension 3000, EasyScan2, and ezAFM should have the signal-to-noise ratio to measure individual DNA molecules.

To check if the instruments were able to visualize DNA, we scanned a DNA standard (K-Tek, DNA01) that had individual 1009-nm-long DNA molecules affixed to a mica substrate (Fig. 3). The nominal diameter of a double-stranded DNA molecule is 2 nm, but when DNA is adhered to mica using a salt solution, the DNA is embedded into a salt layer, which distorts the height of the molecule.[17] Typically, measurements of DNA height above the salt layer are around 0.5 nm.[17] The Dimension 3000 and the ezAFM were both able to resolve the molecules and recorded heights of 0.55 ± 0.07 nm and 0.32 ± 0.08 nm, respectively. Measurements are the mean ± standard deviation for the Gaussian fits to the peaks in Fig. 3. The lower height recorded by the ezAFM probably indicates that the probe is distorting the molecule. The lateral width measured by each AFM should be a convolution of the width of the feature and the width of the AFM probe (Fig. 1, *dashed line*). Thus, the expected lateral radius of the feature is estimated to be $\sqrt{4rR}$, where $r$ is the actual radius of the feature and $R$ is the radius of the probe.[17] DNA strands imaged by the Dimension 3000 had a radius of 3.9 ± 0.6 nm, close to the expected lateral radius of 3 nm. DNA strands imaged by the ezAFM had a radius of 8.1 ± 0.5 nm, a factor of two higher than the expected lateral radius of 4.5 nm. This larger measurement is most likely due to a dull tip or an underestimate of the nominal tip radius by the manufacturer.

When we scanned the DNA standard with the EasyScan2, we were not able to resolve individual DNA molecules, even though the noise level of the instrument (0.32 ± 0.09 nm) was theoretically low enough. The reason for this inability is most likely due to the fact that the instrument is operating in contact mode. In contact mode, there is a lateral force on the sample which can distort or dislodge soft biological molecules as the AFM tip scans across the surface.

Perhaps tip distortion at the level of a few Ångstroms or the lack of adhesion of the molecules to the surface prevents us from imaging the DNA. In any event, we note that we were able to resolve larger biological molecules (microtubules and molecular clumps of λ phage DNA) with the EasyScan2 and cells with the TKAFM, indicating that contact mode AFMs are still a useful instrument for the biophysics laboratory.

In comparing these different instruments, we find a wide range of resolutions and capabilities. Selection of the right AFM for your classroom will depend on your needs. The Dimension 3000 has the flexibility and atomic-scale resolution of a research instrument and is available for use at many AFM user facilities, but purchasing an instrument is fairly expensive and running the instrument requires some training. The EasyScan2 and ezAFM are less expensive than the Dimension 3000 and both are portable, user friendly, and have atomic-scale resolution. However, they require particular cantilevers and are not as flexible as the other two instruments. Finally, the TKAFM allows students to build their own AFM during the laboratory, and as such is the least expensive AFM we tested with the most limited resolution. We note that in reviewing these instruments, our goal here is not to advocate for one type of AFM over another, but to create biophysics laboratories for the range of educational AFMs that exist.

## IV.     MEASURING YEAST CELL MORPHOLOGY TO DETERMINE MEMBRANE TENSION

For AFMs with a high noise floor (>20 nm) or for AFMs in contact mode, a good biophysical experiment is to image the morphology of *S. cerevisiae* (baker's yeast) (Fig. 4). Yeast cells have a cell wall that protects them from the high forces (nanoNewtons) that might be encountered during imaging and are large enough (~3 μm in diameter) to be easily imaged. In addition, yeast cells are visible using bright field microscopy, so students can check the viability of samples ahead of time or could make complementary dynamic measurements of yeast growth[29] using traditional microscopy.

Here, our goal is to measure the morphology of budding yeast cells and determine the relative tension in the membrane between the mother and bud. Previously,[30] membrane tension for yeast has been estimated by using the Young-Laplace equation, which determines the pressure difference, $\Delta P$, between the inside and outside of the cell based on the local surface tension, $\gamma$, and the shape of the cell:

$$\Delta P = \gamma \left(\frac{1}{R_1} + \frac{1}{R_2}\right). \tag{1}$$

Here $R_1$ and $R_2$ are the semi-major and semi-minor axes, respectively, of the ellipsoidal cell. Students can derive this equation by setting the tensional force in the membrane equal to the force perpendicular to the membrane that causes the pressure difference. Since the pressure difference between the inside and outside of the cell should be the same everywhere due to Pascal's principle, the surface tension in the membrane will solely depend on the shape of the cell.

To perform the experiment, we have the students image yeast cells from two different samples: one that is saturated with cells and therefore has cells without buds (Fig. 4A), and one that has yeast cells that are still in the exponential growth phase and therefore has cells with buds (Fig. 4B). Images are taken with the TKAFM. To analyze the data, the students identify a yeast cell with a bud and fit two ellipses by eye: one to the mother cell and one to the bud, recording the semi-major and semi-minor axes of each ($R_{1m}$, $R_{2m}$, $R_{1b}$, and $R_{2b}$, respectively). This allows them to estimate the relative surface tension in the bud by calculating the ratio:

$$\frac{\gamma_b}{\gamma_m} = \left(\frac{1}{R_{1m}} + \frac{1}{R_{2m}}\right) / \left(\frac{1}{R_{1b}} + \frac{1}{R_{2b}}\right), \tag{2}$$

where $\gamma_m$ and $\gamma_b$ are surface tensions in the mother and bud, respectively.

The students should find that the bud is never bigger than the mother cell, a common occurrence in budding organisms,[31] and therefore the measured ratio should indicate less surface tension in the bud. The ratio we measure for tension in the mother to tension in the bud (Fig. 4) is 0.7. Some organisms (like *Clostridium tetani*, the bacteria that causes tetanus) have offspring (spores) that are larger than the mother cell, and many other bacteria divide by binary transverse fission producing two identically sized cells.[31] Interestingly, the growth machinery for yeast is inserted into areas of the membrane with less tension, possibly setting the physical limits of the shape of the yeast cell.[30]

## V.  MEASURING THE EFFECT OF GLUTARALDEHYDE ON MICROTUBULE PERSISTENCE LENGTH

A great biophysical experiment for AFMs with a lower noise floor (<20 nm) or for contact mode AFMs is to measure the persistence length of a microtubule adhered to the surface. In polymer physics, the persistence length, $L_p$, is used to describe the flexibility of a polymer, and is essentially the average length of a straight section along the polymer chain. For example, the persistence length for uncooked spaghetti is larger than a meter, and therefore spaghetti in a box will appear as a rigid rod. Cooked spaghetti, on the other hand, might have a persistence length of a centimeter, and will appear as a flexible coil on your plate. Mathematically, the persistence length is the decay length for the slope of a tangent vector to the polymer, which decays according to $e^{-s/L_p}$.[32] Here the variable $s$ is the distance along the polymer.

Having a good understanding of persistence length is important when thinking about the mechanics of the cell. There are three main types of biological filaments in a cell: microtubules, actin, and intermediate filaments. Intermediate filaments are the most flexible with a persistence length of order 1 μm,[33] while actin and microtubules have a persistence length of order 10 μm and 1 mm, respectively.[34] The cell can change the stiffness of a particular region by cross-linking filaments or by changing the concentration of filaments.[35] Here we will change the apparent stiffness of a microtubule by adding the fixing agent glutaraldehyde.[36]

In this experiment, we take an image of two samples of microtubules adhered to a mica substrate with the EasyScan2 (Fig. 5) and compare the apparent persistence length of each sample. One sample has a glutaraldehyde concentration of 1%, and the other has a glutaraldehyde concentration of 4%. To estimate the persistence length, we measure the length of a "straight" section of a microtubule in the image by eye. After measuring 10 such lengths, we calculate the average and standard deviation as our estimate of the persistence length. For the 1% glutaraldehyde sample, we find a persistence length of 16 ± 8 μm in an image that is 50 by 50 μm. For the 4% glutaraldehyde sample, we measured a persistence length of 1 ± 0.6 μm for the 10 by 10 μm image in Fig.5. When we compare the two measurements we find that the glutaraldehyde decreases the apparent persistence length of the molecule.

This result is often not intuitive for students as many hypothesize that a fixative would act to suppress molecular movements and decrease the flexibility of the polymer. Given this hypothesis, students generally predict that the persistence length would increase upon increasing the glutaraldehyde concentration. However, our images show the exact opposite result. It turns out that the microtubule distorts upon cross-linking with the aldehyde, decreasing the apparent persistence length,[36] instead of increasing it.

Students also notice that our measured persistence length for microtubules in the 1% glutaraldehyde sample is two orders of magnitude below the persistence length for a freely mobile microtubule in water.[34] Part of this is attributable to the glutaraldehyde and part of this is attributable to the adsorption to the surface. The adsorption process tends to constrain the molecule in both the 1% and 4% glutaraldehyde samples, lowering the apparent persistence length for both cases.[33] In addition, since many of the microtubules in the 1% glutaraldehyde sample leave the field of view before bending, our measured persistence length may be artificially low. Images with larger fields of view may not be possible as the field of view is limited by the scan range for the instrument.

## VI.    CALCULATING THE CONTOUR LENGTH OF DNA

If your AFM has an even lower noise floor (~0.1 nm) and operates in tapping mode, then you can perform some particularly rich experiments imaging DNA. In these experiments, you will be able to resolve the molecular conformation of the DNA and can see the random walk of the polymer, allowing for measurements of contour length.

In statistical mechanics, the random walk is generally described by a one-dimensional stepper starting at the origin and taking a step with distance ($\delta$) either left or right, randomly. The mean displacement ($\bar{x}$) of an ensemble of these one-dimensional steppers is zero, however the one-dimensional steppers do start to spread away from the origin after a while. If you square the displacement and then take the mean of the ensemble ($\overline{x^2}$), you can prove that this value is nonzero. In fact, this mean squared displacement increases with the number of steps ($N$):

$$\overline{x^2} = N\delta^2. \tag{3}$$

Students can do this proof by finding the mean squared displacement for one step, two steps, three steps, and then generalizing to $N$ steps. The equation they derive, Eq. 3, is just Einstein's 1905 equation for particle movement due to Brownian motion,[37] rewritten in terms of the number of steps. Interestingly, the root-mean-squared (rms) distance of a random walker is proportional to the square root of the number of steps, instead of the number of steps.

Here, we will assume that the molecular conformation of DNA can be described by a random walk. This random walk model is called the Freely Jointed Chain (FJC) model,[32] since we assume that DNA is made up of a series of segments that are linked together in a chain with freely rotating joints. The length of each segment will be given by twice the persistence length,[32] and the number of steps will be given by the total length of the polymer divided by the length of a segment. This total length of the polymer from one end along the contour of the molecule to the other end is called the contour length ($L_c$), and is different from the shortest distance between the two ends, called the end-to-end distance ($R$). In this FJC model, the mean squared end-to-end distance of the molecule is

$$\overline{R^2} = 2L_c L_p, \qquad (4)$$

which can be derived from Eq. 3.[38] Using this equation, we need only measure the persistence length and the rms end-to-end distance of the DNA to find the contour length.

However, it is hard to find the ends of the DNA molecule in an AFM image, and so instead of measuring the end-to-end distance, a better parameter is the radius of gyration of the molecule. The radius of gyration ($R_G$) is just the rms distance from the origin to all of the segments in the chain. For molecules where $L_c \gg L_p$, the radius of gyration is related to the end-to-end distance by:

$$\overline{R_G^2} \approx \frac{\overline{R^2}}{6} \approx \frac{L_c L_p}{3}.^{38} \qquad (5)$$

If we determine the persistence length and radius of gyration from an image of a DNA molecule, we can calculate the contour length of the molecule.

Here, we will image either double-stranded or single-stranded DNA with the Dimension 3000 and estimate the radius of gyration and persistence length to find the contour length of the DNA. First, we image a double-stranded, λ phage DNA molecule (Fig. 6A). To estimate the radius of gyration for this molecule, we determine the diameter of the molecule in two dimensions ($D_1$ and $D_2$, respectively), average the two diameters, and divide by two. Measurements of multiple molecules ($N=5$) produce an average radius of gyration of 0.49 ± 0.19 μm. We then zoom in on the double-stranded DNA (Fig. 6B) and measure the persistence length to be 55 ± 16 nm using the procedure in the previous section. Interestingly, this measurement of the persistence length agrees with single molecule stretching experiments[32] and indicates that our surface preparation did not distort the molecule. Finally, we estimate the contour length of the molecule as 13 ± 12 μm using Eq. 5, which agrees with the nominal contour length of 16.4 μm.

Using a similar method, we estimate the contour length of single-stranded DNA (Fig. 6C) to be 1900 nm ± 500 nm. In this method, we measure the radius of gyration to be 36 ± 4 nm (*N*=8) and then use the known persistence length for single-stranded DNA of 2 nm.[39] This estimate is within a factor of 1.3 of the nominal value of 2450 nm.

## VII. IMAGING DNA UNDER TENSION AND DETERMINING THE STRETCHING FORCE

Another interesting experiment with DNA that does not require as much resolution (only ~1 nm) is to image double-stranded DNA that has been adhered to a surface while under the tension of a flow. The force of the flow stretches the DNA, which behaves as an entropic spring at low force. By measuring the extension of the DNA in the AFM image, we can estimate the force of the flow.

If we assume DNA behaves as a freely jointed chain, then as we stretch the DNA, we bias the random walk of the molecule in a particular direction. This means that as the force increases, there is a decrease in the number of conformations or states that the molecule can access, and we get an overall decrease in the entropy. From statistical mechanics, we know that the Helmholtz free energy (*F*) is related to the internal energy (*U*), the temperature (*T*), and the entropy (*S*) through the equation:

$$F = U - TS. \tag{6}$$

If there is a decrease in the entropy, this creates an increase in the free energy much like stretching a spring creates an increase in the potential energy.

To model the stretching force, we therefore assume that DNA behaves as an entropic spring. To first order, the stretching force on the DNA ($f_{DNA}$) will follow Hooke's Law:

$$f_{DNA} \approx k x_{DNA}. \tag{7}$$

Here, $x_{DNA}$ is the extension of the DNA and *k* is the spring constant. Since *k* should only depend on the thermal energy ($k_B T = 4.1$ pN·nm), the persistence length, and the contour length, we can use unit analysis to guess the form:

$$k \propto \frac{k_B T}{L_p L_c}. \tag{8}$$

Combining Eqns. 7 and 8 we derive the approximate stretching force to be:

$$f_{DNA} \approx \frac{k_B T}{L_p L_c} x_{DNA}. \tag{9}$$

Interestingly, this rough approximation is valid at low forces (<0.1 pN) where entropic stretching dominates. A more rigorous solution for the FJC model can be derived by upper level physics students,[38] however this solution does not include the enthalpic contributions due to bending and stretching the polymer. The inextensible worm-like chain (WLC) model takes into account these contributions and models the polymer as a continuously flexible chain.[40] The interpolation formula for the WLC is:

$$f_{DNA} = \frac{k_B T}{L_p}\left[\frac{1}{4(1-x_{DNA}/L_c)^2} - \frac{1}{4} + \frac{x_{DNA}}{L_c}\right].\text{[40]} \tag{10}$$

Here we see that the linear term in Eq. 10 matches our approximation (Eq. 9). However, we have incorrectly assumed that there is not a zeroth order term. In the FJC model, we would not predict a zeroth order term since the average end-to-end distance for a molecule undergoing a random walk would be zero. The fact that DNA has a nonzero average end-to-end distance means that there are internal interactions within the molecule that we have not accounted for in our simple FJC model.

To perform the experiment, we image double-stranded λ phage DNA molecules adhered to mica in the presence of a flow (Fig. 7) with the ezAFM. We then estimate $x_{DNA}$ by measuring the length of a DNA molecule along the direction of the flow. We repeat this measurement multiple times and find that the average and standard deviation for 5 measurements is $4 \pm 2$ μm, suggesting that the force of the flow is $0.04 \pm 0.03$ pN when using Eq. 10.

Students are usually surprised by this result. Typically, we ask students to estimate the stretching force using Stoke's law before performing the experiment. Stoke's law predicts the drag force on an object with a radius, $a$, moving with velocity, $v$, in a medium with viscosity, $\eta$. This drag force should be equal to the stretching force, giving us the equation,

$$f_{DNA} = 6\pi\eta a v. \tag{11}$$

Here the relative velocity between the object and the medium is the velocity of the flow since we assume that part of the molecule will adhere to the mica while the rest is under the tension of the flow. Many students estimate the flow as about 1 cm/s by watching a drop of water move across the mica while making the sample. The radius of the DNA can be estimated by the radius of gyration of the molecule, which for our λ phage DNA is 0.5 μm in the absence of flow. Since the viscosity of water is 0.001 Pa-s, the drag force should then be 100 pN, a value that is 4 orders of magnitude off from our measurement. Students are stumped, where did they go wrong?

It turns out, that their estimate of the velocity of the flow is incorrect. Fluid flow for liquid on a mica coverslip is similar to fluid flow in a pipe. The velocity of liquid close to the surface is almost zero. To experimentally verify this result, students can flow micron-sized polystyrene particles through a sample chamber[41] and use a light microscope to record their movement. Tens of microns from the coverslip the liquid moves so rapidly it is hard to measure the movement of the bead across the screen by eye, yet close to the coverslip, the velocity of the particles is on the order of 1-10 microns/second, suggesting that our measured force of 0.04 pN is correct.

## VIII.  CONCLUSION

Many different types of AFMs are now available for educational use in the classroom. Here we compare the different instruments for easy selection. In addition, we offer a slate of experiments over a range of AFM resolutions that allow students to both image biological materials (cells, microtubules, and DNA) and measure biophysical parameters (membrane tension, persistence length, contour length, and the stretching force). The described experiments take advantage of the ability of the AFM to measure a wide variety of surfaces and give students the opportunity to image features with a completely different kind of microscope. In the future, these experiments can be improved if the software for the instruments is updated to allow for force-spectroscopy. In that event, students will be able to measure the "important" forces between the AFM probe and the sample, as Binnig, Quate, and Gerber intended.[4]

**ACKNOWLEDGMENTS**

We would like to thank Jenny Ross at the University of Massachusetts and Keisuke Hasegawa at Grinnell for microtubules. We also would like to thank Sekar Thirunavukkarasu at the University of Massachusetts for providing training and experimental advice on the Dimension 3000. This work was supported by Amherst College and Bridgewater State University, including Edith Schoolman funds.

_ _ _ _

[†]arcarter@amherst.edu

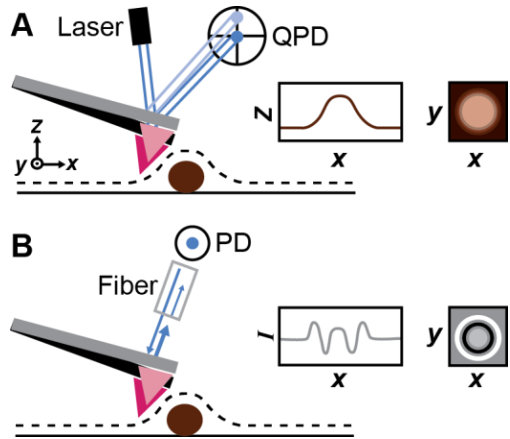

FIG. 1. (Color online only.) Diagram showing the operation of an atomic force microscope (AFM). (A) In a typical AFM, a sharp tip (radius of ~10 nm) at the end of a cantilever that is hundreds of microns long scans the surface. Upon encountering a feature at a different height than the surface, the cantilever tilts and deflects a laser beam. This laser deflection is detected by a quadrant photodiode (QPD) that records the corresponding movement of the cantilever as a change in height ($z$). Multiple scans in $x$ with interspersed movements in $y$ create an image of the feature in $x$ and $y$. (B) In the TKAFM, the cantilever deflection is detected by a photodiode (PD) that monitors the reflected laser light from both the cantilever (thick arrow) and the optical fiber (thin arrow). Interference between the two sources of reflected light cause the intensity on the PD, $I$, to change as the cantilever-fiber distance changes. The number of fringes then gives the height of the feature. Diagrams not to scale.

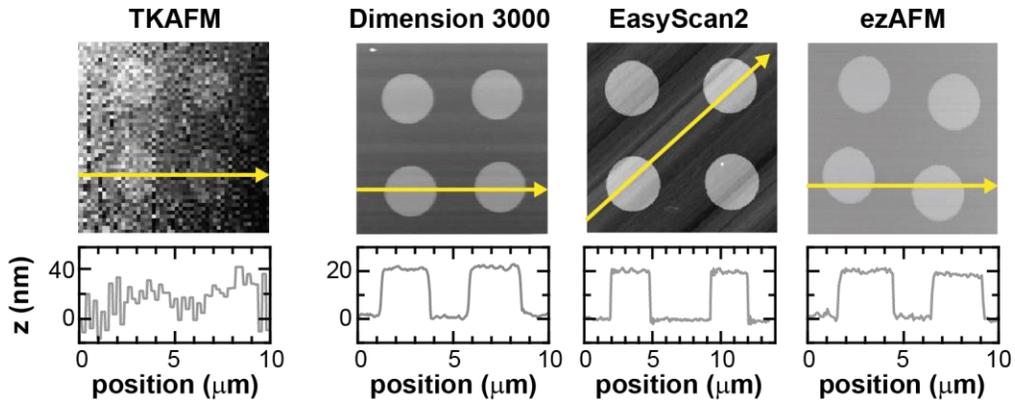

FIG. 2. (Color online only.) Images of a commercial AFM height standard show the signal-to-noise ratio of the different instruments. The instruments are the TKAFM from Thorlabs, the Dimension 3000 from Digital Instruments, the EasyScan2 from Nanosurf, and the ezAFM from Nanomagnetics. The height standard contains 20-nm-tall silicon pillars that are 5 μm in pitch. The arrow through each image marks the location of the line scan shown below the image. Line scans depict the height ($z$) of the pillars above the surface and are used as an estimate of the signal-to-noise ratio.

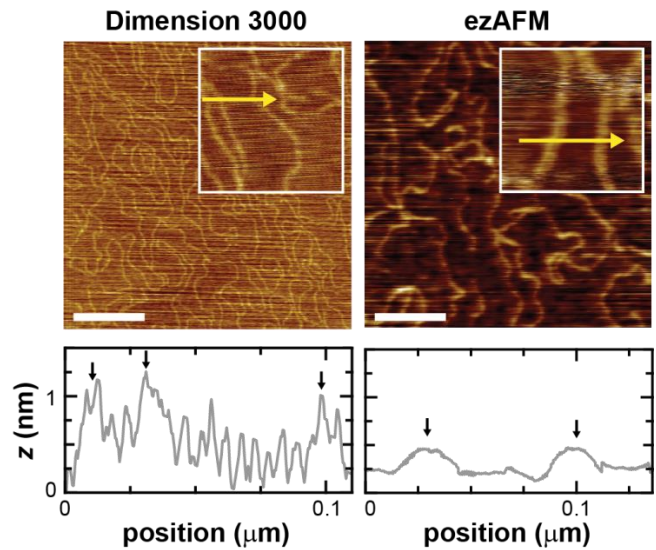

FIG. 3. (Color online only.) Images of a commercial DNA standard allow for comparison of the tapping mode instruments. The DNA standard has ~10-500 molecules/μm$^2$ at a height of 0.5 nm. The region imaged with the ezAFM is less concentrated and the radius of the AFM probe on the ezAFM is larger. Insets have a width of 200 nm and show a zoom in of the DNA. Profiles depict the height ($z$) of the features along the arrow drawn in the image. Black arrows in the profile note the locations of the DNA molecules. Scale bars are 250 nm.

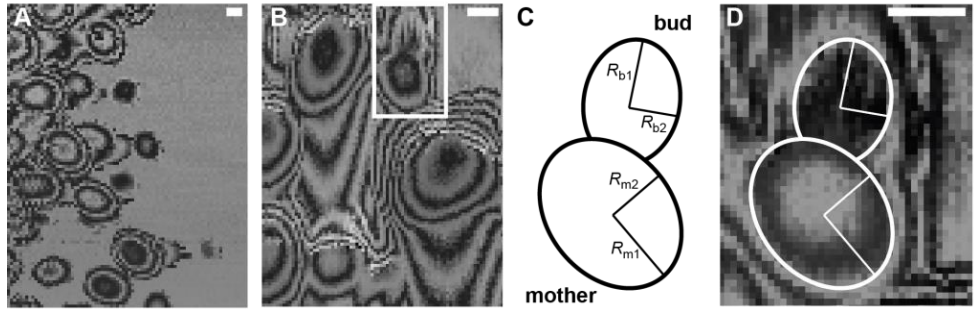

FIG. 4. Images taken with the TKAFM allow for measurements of yeast cell morphology. (A) Image of yeast cells on a glass surface depict the ellipsoidal nature of the strain. Height above the surface is given by the number of fringes. Cells are about 3 µm tall. (B) Image of yeast cells adhered to the glass during an exponential growth phase. Some cells have buds (*white box*). (C) The cell morphology of the mother and the bud can be analyzed by fitting an ellipse to each and finding the semi-major and semi-minor axes ($R_{1m}$, $R_{2m}$, $R_{1b}$, and $R_{2b}$, respectively). (D) Another image of the cell in the white box in *B* taken at a higher magnification. Ellipses are fit to the data by eye. Scale bars are 1 µm.

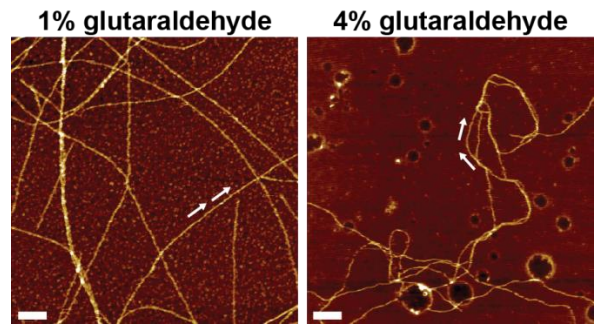

FIG. 5. (Color online only.) Images taken with the EasyScan2 of microtubules in the presence of either 1% or 4% glutaraldehyde. The length over which the tangent vector to the microtubule (arrows) remains correlated is greater in the 1% glutaraldehyde case, indicating a longer persistence length. Scale bars are 1 μm.

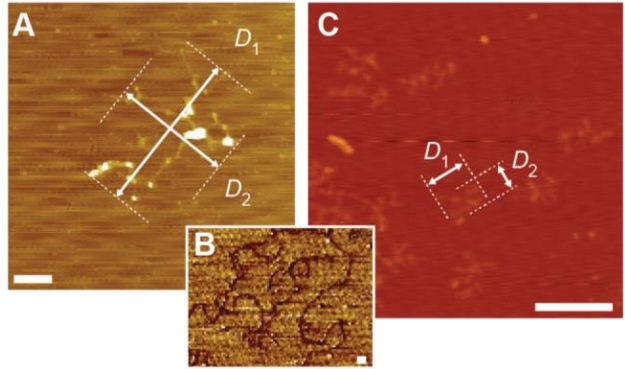

FIG. 6. (Color online only.) Images of double-stranded (A-B) and single-stranded (C) DNA molecules taken with the Dimension 3000 show the random walk of the polymer. Measurement of the radius of gyration (estimated as half the average of $D_1$ and $D_2$) allows for calculation of the contour length. The persistence length of the double-stranded DNA can also be estimated from a higher resolution phase image shown in *B*. Scale bars are either 200 nm (*A* and *C*) or 50 nm (*B*).

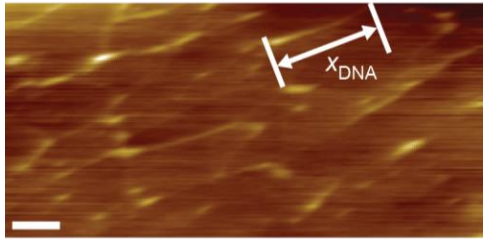

FIG. 7. (Color online only.) Images of λ phage DNA stretched in the direction of a flow to a particular extension ($x_{DNA}$). Measuring the extension allows students to determine the force of the flow. Scale bar is 2 μm.

| Figure | Sample | AFM | Tip | Scan Rate | Original Image Size | P-I-D | Set Point |
|---|---|---|---|---|---|---|---|
| 2 | HS-20mg | TKAFM | thorlabs interdigitated cantilever | 20 µm/s | 20 x 20 µm (100 x 100 pixels) | - | - |
| 2 | HS-20mg | Dimension 3000 | PPP-NCLR | 40 µm/s | 20 x 20 µm (512 x 512 pixels) | 0.36-0.16- | 1.605 V |
| 2 | HS-20mg | EasyScan2 | PPP-XYNCSTR | 20 µm/s | 20 x 20 µm (512 x 512 pixels) | 10000-1000- | 20 nN |
| 2 | HS-20mg | ezAFM | PPP-NCLR | 5 µm/s | 20 x 20 µm (256 x 256 pixels) | 80%-1%-60% | 50% |
| 3 | DNA01 | Dimension 3000 | hires19 | 2 µm/s | 1 x 1 µm (512 x 512 pixels) | 0.36-0.16- | 1.55 V |
| 3 inset | DNA01 | Dimension 3000 | hires19 | 2 µm/s | 200 x 200 nm (512 x 512 pixels) | 0.36-0.16- | 1.55 V |
| 3 | DNA01 | ezAFM | PPP-XYNCSTR | 1 µm/s | 1 x 1 µm (512 x 512 pixels) | 10%-1%-7% | 70% |
| 3 inset | DNA01 | ezAFM | PPP-XYNCSTR | 200 nm/s | 200 x 200 nm (512 x 512 pixels) | 10%-1%-6.4% | 50% |
| 4A | yeast cells from saturated plate | TKAFM | thorlabs interdigitated cantilever | 10 µm/s | 20 x 20 µm (100 x 100 pixels) | - | - |
| 4B | yeast cells in growth phase | TKAFM | thorlabs interdigitated cantilever | 10 µm/s | 10 x 10 µm (100 x 100 pixels) | - | - |
| 4D | yeast cells in growth phase | TKAFM | thorlabs interdigitated cantilever | 10 µm/s | 8 x 8 µm (100 x 100 pixels) | - | - |
| 5 | microtubules in 1% glutaraldehyde | EasyScan2 | PPP-XYNCSTR | 10 µm/s | 10 x 10 µm (512 x 512 pixels) | 10000-1000- | 20 nN |
| 5 | microtubules in 4% glutaraldehyde | EasyScan2 | PPP-XYNCSTR | 10 µm/s | 10 x 10 µm (512 x 512 pixels) | 10000-1000- | 20 nN |
| 6A | Lambda DNA | Dimension 3000 | PPP-XYNCSTR | 3 µm/s | 3 x 3 µm (512 x 512 pixels) | 0.36-0.288- | 1.7 V |
| 6B | Lambda DNA | Dimension 3000 | PPP-XYNCSTR | 10 µm/s | 5 x 5 µm (512 x 512 pixels) | 0.36-0.16- | 1.5 V |
| 6C | M13mp18 single-stranded DNA | Dimension 3000 | PPP-XYNCSTR | 1 µm/s | 1 x 1 µm (512 x 512 pixels) | 0.36-0.16- | 1.5 V |
| 7 | Lambda DNA with flow | ezAFM | SSS-FMR | 20 µm/s | 20 x 20 µm (1024 x 1024 pixels) | 20%-1%-35% | 60% |

TABLE 1. Experimental parameters for all images.